\documentclass[aps,prb,showpacs,floatfix,superscriptaddress,twocolumn,byrevtex]{revtex4-1}
\pdfoutput=1
\usepackage{amsmath}
\usepackage{amssymb}
\usepackage{amstext}
\usepackage{amsopn}
\usepackage{amsfonts}
\usepackage{amsxtra}
\usepackage{color}
\usepackage{dcolumn}
\usepackage{graphicx}
\usepackage{bm}
\usepackage{hyperref}

\begin{document}

\title{Bi$_2$Te$_{3-x}$Se$_x$ series studied by resistivity and thermopower}

\author{Ana Akrap}
\email{ana.akrap@unige.ch}
\author{Alberto Ubaldini}
\author{Enrico Giannini}

\affiliation{University of Geneva, CH-1211 Geneva 4, Switzerland}
\author{L\'aszl\'o Forr\'o}
\affiliation{EPFL, Lausanne, Switzerland}
\date{\today}

\begin{abstract}
We study the detailed temperature and composition dependence of the resistivity, $\rho(T)$, and thermopower, $S(T)$, for a series of layered bismuth chalcogenides Bi$_2$Te$_{3-x}$Se$_x$, and report the stoichiometry dependence of the optical band gap. In the resistivity of the most compensated member, Bi$_2$Te$_{2.1}$Se$_{0.9}$, we find a low-temperature plateau whose onset temperature correlates with the high-temperature activation energy. For the whole series $S(T)$ can be described by a simple model for an extrinsic semiconductor. By substituting Se for Te, the Fermi level is tuned from the valence band into the conduction band. The maximum values of $S(T)$, bulk band gap as well the activation energy in the resistivity are found for $x \approx 0.9$.
\end{abstract}

%
%  PACS numbers
%

\maketitle

Bi$_2$Te$_{3-x}$Se$_x$ are a family of narrow-band semiconductors, well known since the 1950's for their exceptional thermoelectric properties.\cite{goldsmid64} 
Bi$_2$Te$_3$ has the highest known thermoelectric figure of merit $zT$ at room temperature, and is widely used in room-temperature thermoelectric applications. 
Recently, bismuth chalcogenides were rediscovered within the novel context of topological insulators.\cite{zhang09,wang11} The surface states in Bi$_2$Te$_{3-x}$Se$_x$ are simple since only one Dirac  cone crosses the band gap.\cite{hasan09,valla12} However, in spite of a reasonably large band gap ($200 - 300$~meV), bulk conductivity of known bismuth chalcogenides is relatively high and hinders the observation of topological surface states, even in the most compensated compound Bi$_2$Te$_2$Se.\cite{ren10,akrap12} Throughout this series, the Fermi level is found either within the conduction or valence band.
In this paper we study the detailed temperature dependence of the resistivity $\rho(T)$ and thermoelectric power $S(T)$ of bismuth chalcogenides to clarify the relation between the composition and position of the Fermi level, and to address the possible role of the surface states in the thermoelectricity of these compounds. We investigate nine different compositions of the Bi$_2$Te$_{3-x}$Se$_x$ series, where Se content is tuned from $x=0$ to 3. The thermopower $S(T)$ may be semi-quantitatively  explained by the standard extrinsic semiconductor model.\cite{goldsmid64}  
$S(T)$ shows that tuning Te/Se content leads to a steady shift of chemical potential, from the valence band in the p-type Bi$_2$Te$_3$, to the conduction band in the n-type Bi$_2$Se$_3$. The highest value of the thermopower is found at $x=0.9$ and it corresponds to the lowest Fermi temperature. 
For this composition, two electronic contributions are needed to describe both $\rho(T)$ and $S(T)$ at low temperatures. While this is consistent with the effect of topological surface states, it is more likely that the charged defects in the material lead to nearly resonant hopping and thus shortcut the conduction at low temperatures.

Single crystals of Bi$_2$Te$_{3-x}$Se$_x$ were grown by the floating zone method from the stoichiometric ratio of metallic bismuth and chalcogenide  elements.
The unit cell of a Bi$_2$Te$_{3-x}$Se$_x$ compound consists of quintuple Te/Se--Bi--Te/Se--Bi--Te/Se layers stacked along the {\em c}-axis direction.\cite{nakajima63,bayliss91} The quintuple layers are bound by weak van der Waals interaction. We investigated nine different compositions, given by the Se content $x=0,0.6,0.9,0.95,1,1.3,1.5,2$ and 3. Crystals were cleaved and cut into thin bars of approximately $1.5 \times 0.5 \times 0.02$~mm$^3$. For the resistivity measurements of the $x=0.9$ composition, samples were cleaved to seven different thicknesses, from 4~$\mu$m to 140~$\mu$m. The resistivity was measured using a standard four-probe technique. Thermopower was determined simultaneously using a $dc$ method, where the sample was fixed to a ceramic surface with a small heater attached next to it, and the thermal gradient was determined using a chromel-constantan differential thermocouple. 
The bulk band gap was determined through transmission and reflection measurements on cleaved thin flakes, using a Bruker Hyperion microscope, by following the energy of the onset of the Fabry-Perot interferences, similar to the previous high pressure studies of bismuth chalcogenides.\cite{vilaplana11a,segura12,akrap12} 

%
% Figure 1
%
\begin{figure*}[htb]
\centering
\includegraphics*[width=\textwidth]{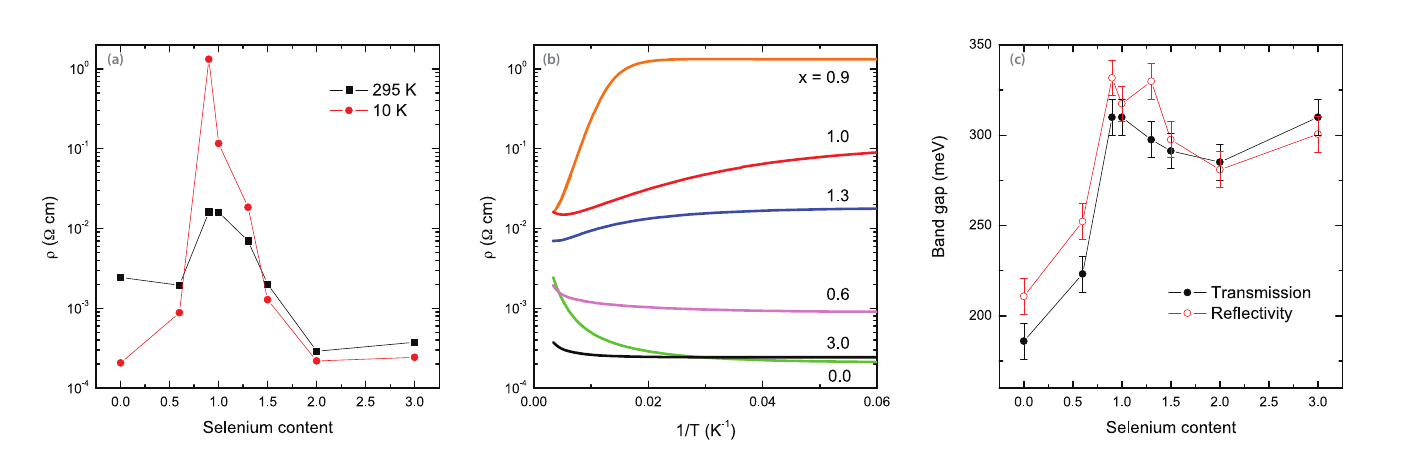}
\caption{(a) Resistivity at high and low temperature for varying composition, from Bi$_2$Te$_3$ to Bi$_2$Se$_3$. A maximum $\rho$ occurs for $x=0.9$. (b) Temperature dependence of resistivity for several different compositions $x$. (c) The dependence of bulk band gap on Selenium content $x$.
\label{fig:rho_xT}}
\end{figure*}

The behavior of the resistivity throughout the series is shown in Fig.~\ref{fig:rho_xT}(a) at room temperature and at 10~K. At room temperature a pronounced maximum in $\rho$ occurs for $x\sim1$. At low temperature it becomes clear that the resistivity is the largest for $x=0.9$. A more detailed temperature dependence of the resistivity is shown in Fig.~\ref{fig:rho_xT}(b) in an Arrhenius plot, for several different compositions $x$. While most samples exhibit a weakly metallic resistivity,  we observe an activated behavior for $x=0.9,0.95,1$ and 1.3, in the $x$ range where $\rho(x)$ has maximum value; this is in agreement with previous work.\cite{jia11,jia12} The activation energies are small and range from $\sim 4.5$~meV (50 K) for $x=0.95, 1$ and 1.3, to  $\sim 47$~meV (540 K) for $x=0.9$. 

The stoichiometry dependence of the bulk band gap is shown in Fig.~\ref{fig:rho_xT}(c). It spans from approximately 200~meV in Bi$_2$Te$_3$ to 300~meV in Bi$_2$Se$_3$, reaching a maximum of 325~meV in Bi$_2$Te$_{2.1}$Se$_{0.9}$. It is interesting to note that the maximum band gap takes place for the same stoichiometry ($x\sim 1$) for which resistivity is maximum. However, even when the resistivity is activated, the transport gap (activation energy) is much smaller than the band gap.\cite{skinner12} While the band gap is determined by the band structure, the transport gap is linked to the presence of charged point defects. When the composition is tuned from Bi$_2$Te$_3$ to Bi$_2$Se$_3$, one passes from a system with excess Te towards a Se-deficient compound, {\em ie} from a hole-rich to an electron-rich system. Around the composition $x=1$, the Se/Te sublattice is expected to be ordered.\cite{jia11}
In addition, the native $p$ and $n$ type defects compensate, decreasing conductivity.\cite{fuccillo13} Therefore around $x=1$ the bulk conductivity is minimum and this composition corresponds to an almost complete compensation of donor and acceptor impurities. 

%
% Figure 2
%
\begin{figure}[htb]%
\includegraphics*[width=\linewidth]{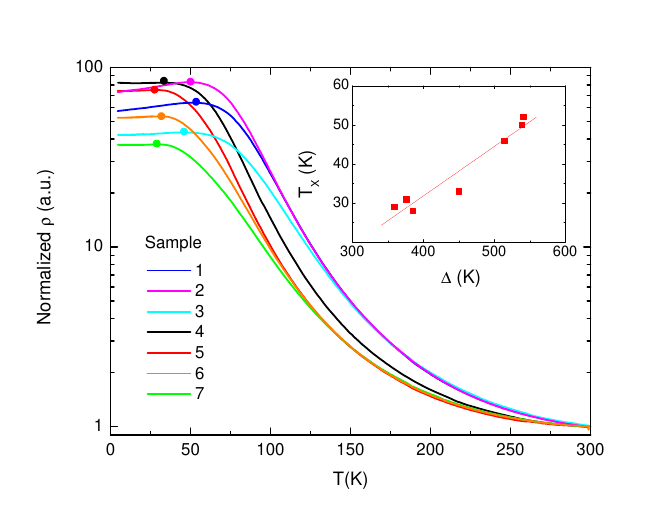}
\caption{The resistivity for $x=0.9$ for seven different samples with nominal $x=0.9$ composition. Filled circles mark the temperature $T_X$ for each curve. Inset: $T_X$ versus the activation energy $\Delta$. Here $T_X$ is determined from the condition that the derivative of the resistivity is zero.
\label{fig:rho09}}
\end{figure}
%

%
% Figure 3
%
\begin{figure*}[htb]%
\includegraphics*[width=\textwidth]{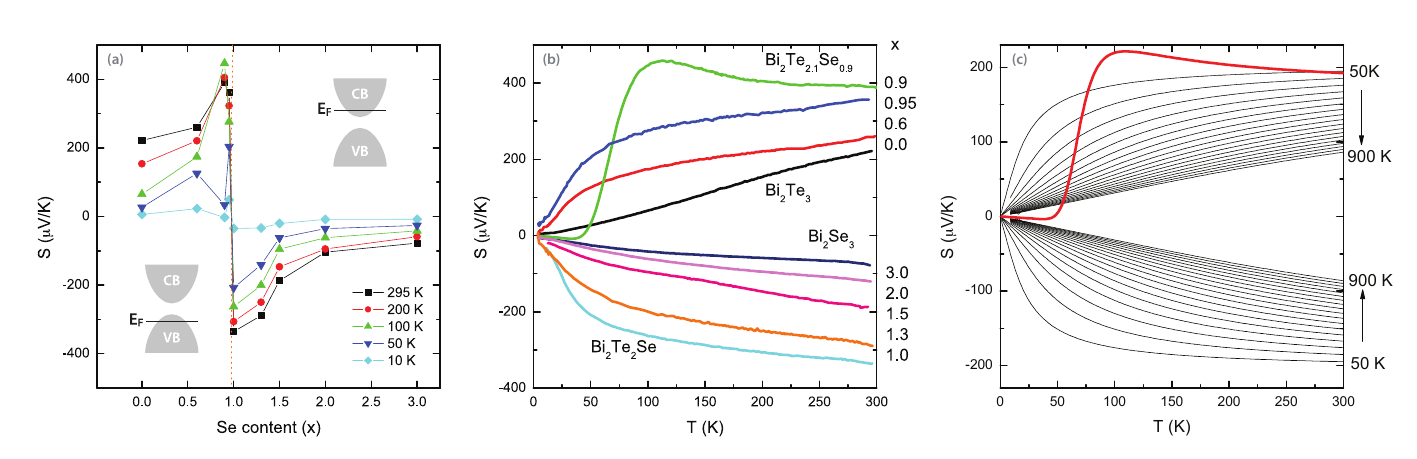}
\caption{(a) The dependence of thermopower on Selenium stoichiometry is shown at five different temperatures.
(b) The measured temperature dependence of $S$ for nine different stoichiometries, given by Selenium content $x$.
(c) Thin black lines are the results of a simple model calculation given by Eq.~\ref{eq:TEP}, where the Fermi level is varied between 50~K and 900~K in steps of 50~K (from 4.3 to 77.6~meV). Thick red line is obtained from Eq.~\ref{eq:TEP_ss}, where the existence of two conducting channels is assumed, as described in the text. To preserve the scale, the latter thermopower (thick red line) was divided by two. Note the different scales in (b) and (c).
\label{fig:TEP}}
\end{figure*}

The $x=0.9$ sample is the most compensated sample in which the resistivity and the activation energy are the highest.
The high-temperature resistivity is well described by activated behavior (see Fig.~\ref{fig:rho_xT}(b)). According to Skinner {\em et al.}\cite{skinner12} the low temperature resistivity should be described by variable range hopping (VRH). However, at low temperatures we do not see any evidence of VRH. Instead, below a temperature $T_X$ a plateau in the resistivity sets in (similar to the previous work\cite{ren10,jia11,jia12}) and $\rho(T)$ starts to saturate or even weakly decrease, as seen in Fig.~\ref{fig:rho_xT}(b), suggesting that another conduction channel becomes dominant at low temperatures. The position of this low-temperature plateau proves to be strongly dependent on the activation energy. To illustrate this, Fig.~\ref{fig:rho09} shows $\rho(T)$ determined for seven $x=0.9$ samples with different dimensions, all prepared from the same starting crystal. To facilitate the comparison, the resistivity curves in Fig.~\ref{fig:rho09} are normalized to 1 at 300~K.

The activation energy $\Delta$ ranges from 360 to 540~K, which shows there are differences in the composition even within the same ingot.\cite{fuccillo13} In the samples with the lowest activation energy, $\Delta < 400$~K, the plateau onset is  at $T_X \sim 30$~K. For the samples with the highest activation energy, $\Delta \approx 540$~K, the plateau starts at $T_X \sim 50$~K. 
If we compare the resistivity of the samples with similar activation energy $\Delta$ (for example, samples 1, 2 and 3 in Fig.~\ref{fig:rho09}), at high temperatures the curves lie on top of each other. However, the resistivity below $T_X$ strongly varies. This may suggest conduction through metallic (topological) surface states,\cite{ren10, shekar14} where the surface states would dominate the conductivity once the bulk resistance becomes sufficiently high. In this case, temperature $T_X$ should be dependent on the sample geometry.
However, a similar resistivity plateau may result from an additional bulk contribution, for example conduction through a weak impurity band in the vicinity of the Fermi level.

Suppose that the bulk conductivity is activated, $\sigma_B=\sigma_0 \exp (-\Delta/T)$. The additional contribution $\sigma_X$ of unknown origin becomes evident at low temperatures and one may ignore its temperature dependence. The total conductivity is then $\sigma_{tot}=\sigma_0 \exp (-\Delta/T) + \sigma_X$.
At high temperatures the bulk conductivity dominates and the total conductivity is activated. At low temperatures the $\sigma_X$ component becomes important. 
If the conduction channel $\sigma_X$ is a bulk contribution like $\sigma_B$, then the geometry of the sample plays no role and the transition temperature will be proportional to the activation energy,
$T_X = \Delta / \ln \left(\frac{\sigma_0}{\sigma_X} \right)$.
The inset in Fig.~\ref{fig:rho09} shows $T_X$ plotted against $\Delta$ for the seven samples from the main panel.The relation is close to linear. Therefore, the low-temperature saturation or decrease in the resistivity is consistent with the formation of an impurity band.\cite{dipietro12, akrap12}

We now come back to the question how the Fermi level position depends on the composition $x$. One way to follow the progression from $p$ to $n$-type conductor is from the thermopower. This is particularly simple in a situation where there is only one relevant electronic band. 
Fig.~\ref{fig:TEP}(a) shows the dependence of thermopower on Se content, $x$, at several different temperatures. There is a very sharp sign change taking place between $x=0.95$ and $x=1.0$. For $x \geq 1$ the thermopower becomes negative, meaning that there is a transition from $p$-type to $n$-type conduction. This transition is illustrated by a cartoon within Fig.~\ref{fig:TEP}(a).

For thermoelectric applications, it is important to optimize the figure of merit $zT$. In applications that do not require maintaining a temperature gradient,  the power factor $S^2/\rho$ should be maximized. The power factor ranges between 10 to 50~$\mu$W/(K$^2$cm) for the series of the samples we study here. These values are similar to those of Sokolov {\emph et al}, and display a similar doping dependence.\cite{sokolov07} Maximum power factor of 50~$\mu$W/(K$^2$cm) takes place for $x=2$, and minimum 10~$\mu$W/(K$^2$cm) occurs for $x=1$ where both the thermopower and the resistivity are maximum.

Figure~\ref{fig:TEP}(b) shows a detailed temperature dependence of thermopower $S(T)$ for nine different compositions $x$. 
All of the samples are either $p$ or $n$ type, and a clear progression can be seen from Bi$_2$Te$_3$ towards Bi$_2$Se$_3$.
While this gradual change from $p$ to $n$ type conductor has been long known, our results also show a very detailed and systematic progression of the temperature dependence of thermopower.
The only exception is the $x=0.9$ sample, which is $p$-type at high temperatures, and $n$-type below $\sim 50$ K. Leaving the $x=0.9$ composition momentarily aside, let us focus on the temperature dependence of thermopower in the remaining eight compounds.
The thermopower of a heavily-doped semiconductor on the metallic side of the metal-insulator transition is given by:\cite{goldsmid64}
\begin{equation}
S=\pm \frac{k_B}{e} \left[ \eta_F - \frac{(r+5/2)F_{r+3/2}(\eta_F)}{(r+3/2)F_{r+1/2}(\eta_F)} \right]  
\label{eq:TEP}
\end{equation}
Here $\eta_F=E_F/(k_B T)$ is the reduced Fermi energy; the parameter $r$ describes the energy dependence of the scattering time, and
\begin{equation}
F_{n}=\int_0^\infty {\rm d} \eta \frac{\eta^n}{1+ \exp \left( \eta - \eta_F \right)}
\label{eq:Fint}
\end{equation}
is the Fermi integral. $E_F$ is measured from the bottom of the conduction band for the $n$-type, or the top of the valence band in the case of a $p$-type semiconductor. To describe the energy dependence of the scattering time parameter, we take the standard value\cite{goldsmid64} $r=-0.5$ for acoustic phonon scattering and calculate $S(T)$ assuming a set of temperature-independent values for $E_F$. 
For Bi$_2$Se$_3$, the carrier density $n$ may be roughly estimated by comparing the resistivity data in Fig.~\ref{fig:rho_xT} to the results reported by Butch {\em et al.},\cite{butch10} which gives a result of $n\sim 10^{18} -10^{19}$ cm$^{-3}$.

Results for the calculated temperature dependence of thermopower are shown in Fig.~\ref{fig:TEP}(c) for $E_F/k_B$, ranging from 50~K to 900~K (corresponding to the $E_F$ between 4.3 and 77.6~meV), in steps of 50~K. This ballpark for $E_F$ seems to be in fair agreement with ARPES results from some members of the series. For example, Hor {\em et al.}\cite{hor09} observed a Fermi level of 50~meV  in Bi$_2$Se$_3$ and Chen {\em et al.}\cite{chen09} saw $E_F\approx45$~meV in Bi$_2$Te$_3$. The measurements of quantum oscillations for Bi$_2$Se$_3$ show that the fundamental state of the system is metallic and that carrier density can be tuned by several orders of magnitude via controlling the stoichiometry.\cite{fauque12} When changing the carrier density from $10^{17}$ to $10^{18}$~cm$^{-3}$, the Fermi temperature moves from 80~K to 1100~K. 
K\"ohler and Hartmann\cite{kohler74} showed that for different carrier densities in Bi$_2$Se$_3$ the energy of the Moss-Burnstein shift in the reflectivity is proportional to the Fermi energy. In the Bi$_2$Se$_3$ sample studied here, the Burnstein peak is at $\sim 0.38$~eV, which gives $E_F \approx 100$~meV. This is in good agreement with the approximate value of 900~K suggested for the Bi$_2$Se$_3$ sample by the calculated series in Fig.~\ref{fig:TEP}(c).

The simple semiconductor model given by the Eq.~\ref{eq:TEP} captures our experimental results quite well. The higher values of $E_F$ lead to a thermopower almost linear in temperature, very similar to the one measured in Bi$_2$Te$_3$ and Bi$_2$Se$_3$. A smaller $E_F$ leads to a thermopower with a distinct change in slope for $k_B T \sim E_F$. Such a shape of $S(T)$ is similar to what we measure for $x\approx 1$. With this in mind, one might conclude from Fig.~\ref{fig:TEP}(b) that samples with $x=0.6, 0.95$, and 1.0 all have $E_F\sim 4$~meV. On the contrary, Fermi levels in pure Bi$_2$Te$_3$ and Bi$_2$Se$_3$ should be higher than room temperature since no clear change in slope is visible from our thermopower data.

While the temperature dependence of $S$ is well-captured by the above model, the absolute values of calculated $S$ are approximately twice smaller than the measured thermopower. This may be due to additional contributions to the thermopower not taken into account by the model, such as phonon drag, or because the parameter $r$ is not simply $-0.5$ (which describes acoustic phonon scattering), but has a temperature and composition dependence.\cite{kulbachinskii12,kulbachinskii97} Chen and Shklovskii\cite{chen13} recently showed that a model of a strongly compensated semiconductor, with shallow and randomly positioned donors and acceptors, gives a good estimate of the absolute value of the thermopower for our most compensated samples, 400~$\mu$V/K.

The thermopower of Bi$_2$Te$_{2.1}$Se$_{0.9}$ has a temperature dependence which differs from the rest of the series. $S(T)$ rises above 400~$\mu$V/K at 100~K, and then precipitously drops below 100~K, changing its sign. This is the only composition for which the sample is $p$-type at high temperatures and $n$-type at low temperatures, suggesting that there is more than one kind of charge carriers. 
Similarly, it was shown\cite{ren10,jia11} that the Hall coefficient of Bi$_2$Te$_2$Se deviates from the activated behavior below 100~K, and changes its sign from positive values at high temperatures to negative values at low temperatures.
More recently, Shekar {\em et al} \cite{shekar14} suggested that the saturation and metallic character of $\rho(T)$ below 30~K was linked to the presence of topological surface states.
%%%%
Among our samples, the $x=0.9$ sample also has the highest resistivity within the series, as shown in Fig.~\ref{fig:rho_xT}(b), and one of the highest bulk band gaps ($\sim 325$~meV). 
Same as in the case of resistivity, we show that this behavior of the thermopower is related to an additional conductivity contribution.
If we suppose there are two parallel channels of conduction in the sample then the total thermopower is given by:
\begin{equation}
S=\frac{S_B \sigma_b+S_X \sigma_X}{\sigma_B+\sigma_X}
\label{eq:TEP_ss}
\end{equation}
Here $S_B$ is the bulk thermopower and $S_X$ is the thermopower of the second conducting channel. 
For a non-degenerate semiconductor, $S_B \propto \frac{k_B}{e} \frac{E_g}{2 k_B T}$ where $E_g$ is the energy gap.\cite{goldsmid64} 
$\sigma_B$ and $\sigma_X$ may be obtained by fitting the resistivity curve for the $x=0.9$ sample shown in Fig.~\ref{fig:rho_xT}(a). For $S_X$ we may suppose either a simple metallic temperature dependence, $S_X=AT$ where $A$ is a constant, or a semiconducting behavior given by Eq.~\ref{eq:TEP}. In both cases, for an appropriate choice of parameters, the total thermopower closely resembles the experimental curve for Bi$_2$Te$_{2.1}$Se$_{0.9}$, capturing the strong decrease of $S(T)$ at 100~K and the sign change between 50 and 100~K as shown in thick red line in Fig.~\ref{fig:TEP}(c).
The experimental $S(T)$ curve for the $x=0.9$ sample agrees fairly well with the above simple calculation, thus confirming the presence of another conduction mechanism in the sample which dominates the resistivity and thermopower at low temperatures.
While this may be due to the surface states, it is likely caused by resonant hopping through the impurity states.

%
% Conclusions
%
To summarize, we have presented a systematic study of the composition and temperature dependence of thermopower within the bismuth chalcogenide Bi$_2$Te$_{3-x}$Se$_x$ series. The maximum resistivity, thermopower, and bulk band gap occur for $x = 0.9$. For this composition, the Te/Se sublattice is presumably maximally ordered which reduces the defects in the crystal lattice.\cite{ren10,jia11} Moreover, at this composition the strong competition between the native $n$ and $p$-type defects can lead to their maximal compensation, and decrease the bulk conductivity.\cite{fuccillo13}
The thermopower at a given temperature can be tuned by changing the stoichiometry. Starting from the $p$-type Bi$_2$Te$_3$, thermopower is strongly enhanced when a part of Te atoms are replaced with Se, and at the composition Bi$_2$Te$_2$Se thermopower changes its sign. As the amount of Se is increased, the value of thermopower decreases. The temperature dependence of thermopower may be semi-quantitatively described by a simple model for an extrinsic semiconductor. Here, the Fermi level $E_F$ decreases from Bi$_2$Te$_3$ towards Bi$_2$Te$_{2.05}$Se$_{0.95}$, reaching a minimum of approximately 4~meV. When more Se is replaced for Te, the Fermi level $E_F$ starts increasing again. Both the bulk band gap and the Fermi level can be tuned by controlling Se/Te stoichiometry. 
For $x=0.9$ we determine the largest absolute values of thermopower and resistivity. At this composition the low temperature electronic transport is dominated by an additional conducting channel.

We would like to thank Kamran Behnia, Brian C. Sales and Nathaniel Miller for helpful advice and comments.
Research was supported by the Swiss NSF through grant No. 200020-135085 and its NCCR MaNEP.
A.A. acknowledges funding from Scholarship of Excellence of the University of Geneva.

\vfil

\bibliography{BiTeSe}

\end{document}